# Reformulating Scalar-Tensor Field Theories

# as Scalar-Scalar Field Theories

# Using a Novel Geometry

by

Gregory W. Horndeski
University of Waterloo
Waterloo, Ontario, Canada
email:
**horndeskimath@gmail.com**

February 22, 2022

# ABSTRACT

abstract
In this paper I shall show how the notions of Finsler geometry can be used to construct a similar geometry using a scalar field, f, on the cotangent bundle of a differentiable manifold M. This will enable me to use the second vertical derivatives of f, along with the differential of a scalar field φ on M, to construct a Lorentzian metric on M that depends upon φ. I refer to a field theory based upon a manifold with such a Lorentzian structure as a scalar-scalar field theory. We shall study such a theory when f is chosen so that the resultant metric on M has the form of a Friedmann-Lemaître-Robertson-Walker metric, and the Lagrangian has a particularly simple form. It will be shown that the scalar-scalar theory determined by this Lagrangian can generate self-inflating universes, which can be pieced together to form multiverses with non-Hausdorff topologies, in which the global time function multifurcates at t=0. Some of the universes in these multiverses begin explosively, and then settle down to a period of much quieter accelerated expansion, which can be followed by a collapse to its original, pre-expansion state.

**Section 1: Cofinsler Spaces and Scalar-Scalar Field Theories**

Modern formulations of classical mechanics employ symplectic structures on cotangent bundles to analyze physical systems. Since general relativity and its alternative formulations of gravity theory can be thought of as a form of classical mechanics one might suspect that the cotangent bundle can also play a basic role in gravity theory. The objective of this paper is to show that is indeed the case for scalar-tensor field theories. I shall demonstrate how that can be done by presenting one way in which Finsler Geometries in the tangent bundle, as described in Finsler [1], Rund [2], Chern, *et al.*, [3], and Pseudo-Finsler geometries, as introduced in Bejancu & Farran [4], can be extended to the cotangent bundle.

Let M be an n-dimensional manifold with cotangent bundle T*M. Recall that $T^*M := \bigcup_{P \in M} T_P^*M$, where $T_P^*M$ is the dual space of $T_PM$, the tangent space to M at P. Let $\pi: T^*M \to M$ denote the canonical (or natural) projection, which maps a covector $\omega \in T_P^*M$ to P: $\pi(\omega) = P$. If x is a chart of M with domain U then it naturally gives rise to a standard chart $(\chi, y)$ of T*M with domain $\pi^{-1}U$ defined by $\chi := x \circ \pi$ and $y(\omega) = y(\omega_i dx^i|_P) := (\omega_1, \ldots, \omega_n) \equiv (\omega_i)$. So if we write y: $\pi^{-1}U \to \mathbb{R}^n$, as $y = (y_1, \ldots, y_n) \equiv (y_i)$, then $y_i(\omega) = \omega_i$, where $\omega = \omega_i \, dx^i|_P$. Now suppose x and x' are overlapping charts of M with domains U and U'. Then $(\chi, y)$ and $(\chi', y')$ are overlapping charts in T*M. On the overlap we have

$$\chi^i = \chi^i(\chi'^j), \quad \chi'^i = \chi'^i(\chi^j) \qquad \text{Eq.1.1a}$$

and
$$y_i = y'_j \frac{\partial \chi'^j}{\partial \chi^i}, \quad y'_i = y_j \frac{\partial \chi^j}{\partial \chi'^i},$$  Eq.1.1b

where the definition of partial differentiation on a manifold was used to make the substitutions

$$\frac{\partial \chi^j}{\partial \chi'^i} = \frac{\partial x^j}{\partial x'^i} \circ \pi \quad \text{and} \quad \frac{\partial \chi'^j}{\partial \chi^i} = \frac{\partial x'^j}{\partial x^i} \circ \pi.$$  Eq.1.2

With these preliminaries disposed of I can define an n-dimensional pseudo-Cofinsler Space. Let f: T*M→ℝ be a smooth function defined on an open submanifold N of T*M where πN = M. Choose a positive integer q such that q<n. If x is any chart of M with domain U, and corresponding standard chart (χ,y) of T*M with domain $\pi^{-1}U$, then we require that ∀ ω∈$\pi^{-1}$U∩N the matrix

$$\left[ \frac{1}{2} \frac{\partial^2 f}{\partial y_i \partial y_j} \right]\bigg|_\omega$$  Eq.1.3

defines a quadratic form on $\mathbb{R}^n$ of index q, which is the number of minus signs when diagonalized. (Due to Eqs.1.1a and 1.1b the index of the matrix presented in Eq.1.3 is independent of the standard chart chosen at ω∈T*M.) When these conditions are met the triple $CF^n$ := (M,N,f), is called an n-dimensional pseudo-Cofinsler Space of index q, and f is called the pseudo-Cofinsler function. When q=1, or n−1, $CF^n$ is called a Lorentzian Cofinsler Space. In what follows our Lorentzian Cofinsler Spaces will have index 1.

It should be noted that in Cofinsler Theory (unlike Finsler Theory) f is not required to satisfy a homogeneity condition.

Suppose we have an n-dimensional pseudo-Cofinsler Space $CF^n = (M,N,f)$ and a scalar function $\varphi$ on M. If the range of $d\varphi$ lies in N; *i.e.,* $d\varphi(M) \subset N$, then we can define the contravariant components of a pseudo-Riemannian metric tensor $g_\varphi$ on M by

$$g_\varphi^{ij} := g_\varphi(dx^i, dx^j) := \tfrac{1}{2}\frac{\partial^2 f}{\partial y_i \partial y_j}(d\varphi), \qquad \text{Eq.1.4}$$

where x is any chart of M with corresponding standard chart $(\chi,y)$ for T*M. Due to Eqs.1.1 and 1.2 it is clear that the $g_\varphi^{ij}$'s are the local contravariant components of a pseudo-Riemannian metric tensor on M.

If one is given an n-dimensional pseudo-Riemannian Space, $V_n = (M,g)$, it is always possible to build a pseudo-Cofinsler Space $CF^n = (M,N,f)$ such that the metric f generates, with any scalar field $\varphi$ on M, is the g of your $V_n$. For just locally define the Cofinser function f by

$$f := (g^{ij} \circ \pi)\, y_i y_j.$$

This Cofinsler function, and corresponding $CF^n$ are said to be trivial.

I shall now present an example of a 4-dimensional Lorentzian Cofinsler Space that will play a significant roll in what follows. To see where this example comes from recall that the FLRW (:=Friedmann [5], Lemaître [6], Roberston [7], Walker [8]) metric in reduced circumference polar coordinates is given by

$$ds^2 = -dt^2 + a(t)^2((1-kr^2)^{-1}dr^2 + r^2d\theta^2 + r^2\sin^2\theta d\zeta^2) \,, \qquad \text{Eq.1.5}$$

where k is a constant with units of (length)$^{-2}$, r has units of length, a(t) is unitless, and c=G=1, with $\theta$ and $\zeta$ denoting the usual polar coordinates on $S^2$. k determines the curvature of the spaces t = constant, which are 3-surfaces of constant curvature. To make matters even simpler I shall for the time being take k=0 in Eq.1.5, and so our line element becomes:

$$ds^2 = -dt^2 + a(t)^2(du^2 + dv^2 + dw^2) \,, \qquad \text{Eq.1.6}$$

where $x \equiv (x^0, x^1, x^2, x^3) =: (t,u,v,w)$, are the standard coordinates of $\mathbb{R}^4$, and the 3-spaces t = constant are flat. What we would like is a function f: $T^*\mathbb{R}^4 \to \mathbb{R}$ which is such that when we construct its associated metric g using Eq.1.4, and a scalar field $\varphi$ on $\mathbb{R}^4$, we get a line element similar to the one presented in Eq.1.6. Well, sure, we could always choose the trivial function $f_T$ defined by

$$f_T := -y_t^2 + a(t \circ \pi)^{-2}((y_u)^2 + (y_v)^2 + (y_w)^2),$$

where $((t,u,v,w) \circ \pi, (y_t, y_u, y_v, y_w))$ is the standard chart of $T^*\mathbb{R}^4$ determined by the chart (t,u,v,w) of $\mathbb{R}^4$. (When working with $T^*\mathbb{R}^4$ I shall adopt the custom of dropping the projection function $\pi$ from the standard chart, since $T^*\mathbb{R}^4$ is obviously diffeomorphic to $\mathbb{R}^4 \times \mathbb{R}^4$.) For this choice of Cofinsler function $d\varphi$ would actually play no roll in the associated metric tensor. A more useful choice of Cofinsler function is

$$f := -y_t^2 + y_t^{-2}((y_u)^2 + (y_v)^2 + (y_w)^2) \,. \qquad \text{Eq.1.7}$$

When $\varphi = \varphi(t)$, this Cofinsler function gives us

$$[g_\varphi{}^{ij}] = \left[ \tfrac{1}{2}\frac{\partial^2 f}{\partial y^i \partial y^j}(d\varphi) \right] = \text{diag}(-1, \varphi'^{-2}, \varphi'^{-2}, \varphi'^{-2}) \qquad \text{Eq.1.8}$$

where a prime denotes a derivative with respect to t. Evidently $g_{ij}dx^i dx^j$ (where here, and in what follows, I shall drop the subscript $\varphi$ on the covariant and contravariant form of the metric tensor) gives us the line element presented in Eq.1.6 with $a(t) = \varphi'(t)$.

At this point some remarks concerning units is in order. We shall, for the most part, use geometrized units in terms of which c=G=1, and all dimensioned quantities will have units of length, $\ell$, to some power. To convey the fact that a physical quantity $\Omega$ has units of $\ell^\lambda$ we write $\Omega \sim \ell^\lambda$. In terms of these units the coordinates t, u, v, w $\sim \ell^1$, and $g_{ij} \sim \ell^0$. Consequently $\varphi' \sim \ell^0$, and hence for us $\varphi \sim \ell^1$. In terms of geometrized coordinates the $y_i$ are unitless. This follows from the fact that

$$y_i(d\varphi) = \frac{\partial \varphi}{\partial x^i} \sim \ell^0.$$

Since $g^{ij} \sim \ell^0$, we can use Eq.1.4 to deduce that $f \sim \ell^0$, as it does in Eq.1.7. Later when dealing with the more general FLRW metric given in Eq.1.5 with $k \neq 0$, I shall still require $g_{ij}dx^i dx^j \sim \ell^2$, $f \sim \ell^0$ and $\varphi \sim \ell^1$, but then some coordinates of T*M will have units and others might be unitless. Lastly, a Lagrange scalar density, L, will be required to have units $\ell^{-2}$, since the integral of L over a 4-volume must have units of action $\sim \ell^2$.

From Eq.1.6, with $a = \varphi'$, we know that the metric on the t=constant slices is

$$(\varphi')^2(du^2 + dv^2 + dw^2) \, .$$

Thus the unitless quantity $|\varphi'|$ plays the roll of a scale factor in these slices. To get the physical distance between any two distinct points in a t=constant slice you just multiply their Euclidean distance by $|\varphi'|$.

The largest subset N of $T^*\mathbb{R}^4$ upon which the function f presented in Eq.1.7 is defined and smooth is N = $\{((t,u,v,w),(y_t,y_u,y_v,y_w)) | \, y_t \neq 0\}$. To determine if f really is a Cofinsler function on N⊂$T^*\mathbb{R}^4$, we need to examine the matrix $g^{ij}$ on N and not just for dφ. This is not a difficult task (*see*, Horndeski [9], noting the error in $g^{tt}$ there), and f is indeed a Lorentzian Cofinsler function on N.

I shall call a Lorentzian Spacetime, $V_4$=(M,g) endowed with a scalar field φ, in which the metric tensor g arises from a Lorentzian Cofinsler Space $CF^4$=(M,N,f) in the manner presented in Eq.1.4, a scalar-scalar theory. So formally a scalar-scalar theory is a pair ($CF^4$=(M,N,f), φ) where dφ⊂N. It is tempting to consider f as a generating function, or as a potential for the Lorentzian metric tensor on M. However, that is not the case since we require both f and φ to generate the metric.

Note that locally there is some "gauge freedom" in f. This is so since we can add an affine function of the form $a^i y_i + b$ to f (where $a^i=a^i(\chi^j)$ and $b=b(\chi^j)$), without effecting the corresponding metric.

The next task is to develop field equations for scalar-scalar theories. This will be done forthwith.

## Section 2: Lagrangians and Field Equations

Suppose that we have a scalar-scalar field theory based on the pair ($CF^4$=(M,N,f), $\varphi$). We would like to build a Lagrangian from f and $\varphi$. The first problem we encounter is that f and $\varphi$ are defined on different spaces. Well, this is easily circumvented by letting $\Phi := \varphi \circ \pi$, where $\pi: T^*M \to M$ is the natural projection. Now f and $\Phi$ are two smooth scalar fields on T*M. Unfortunately it is impossible to use these scalar fields on T*M to build a useful Lagrangian due to the following result proved in [9]:

**Proposition:** If $\mathcal{L}$ is a Lagrange scalar density on an n-dimensional manifold (n>1) which is a concomitant of two scalar field, $\psi_1$ and $\psi_2$, along with their derivatives of arbitrary order, then $\frac{\delta \mathcal{L}}{\delta \psi_1} \equiv 0$ and $\frac{\delta \mathcal{L}}{\delta \psi_2} \equiv 0$. ∎

So let us now consider the possibility of deriving field equations on M using f and $\varphi$. To that end let $\mathcal{L}$ be any second-order Lagrange scalar density on M which is a concomitant of a metric tensor and a scalar field $\varphi$. We replace the metric tensor by the Cofinsler metric tensor built from the vertical derivatives of f evaluated at d$\varphi$, to obtain a Lagrangian of the form

$$\mathcal{L} = \mathcal{L}(g^{ij}(d\varphi); g^{ij}(d\varphi)_{,h}; g^{ij}(d\varphi)_{,hk}; \varphi; \varphi_{,h}; \varphi_{,hk}) \qquad \text{Eq.2.1}$$

where "$_{,h}$" denotes a derivative with respect to the local coordinate $x^h$. $\mathcal{L}$ is locally well-defined on the coordinate domains of M. However, we are now confronted with a problem which is unique to Cofinsler Theory; *viz*., how does one go about computing the variational derivatives of a Lagrangian of the form presented in Eq.2.1? Say we want to vary φ holding f in T*M fixed. Since the $g^{ij}$'s are functions of dφ, any variation of φ will lead to a variation of the $g^{ij}$'s. Conversely, the general variation of $g^{ij}$ would require a variation of both f and φ. In my experience with classical field theories I have not encountered such a convoluted variational problem. Usually when one encounters a Lagrangian which is a concomitant of various fields (all defined on the same manifold), one computes the variational derivative by holding all fields fixed except for one which is varied. That is not possible with the Lagrangian given in Eq.2.1. I shall now describe a method that we can use to circumvent this impediment to computing field equations for scalar-scalar theories. To that end we need to re-examine some very familiar equations from General Relativity.

Let us consider the derivation of the Schwarschild solution [10]. Recall that in this situation we are looking for a static, spherically symmetric, asymptotically flat solution to Einstein's equations $G^{ij} = 0$. The symmetry demands imply that we can choose a coordinate system so that our ansatz metric has the form

$$ds^2 = -e^{\nu}dt^2 + e^{\lambda}dr^2 + r^2d\theta^2 + r^2\sin^2\theta\, d\zeta^2 \qquad \text{Eq.2.2}$$

where ν and λ are functions of r, with θ and ζ being spherical polar coordinates. Conventionally one plugs this metric into $G^{ij} = 0$ to obtain the equations governing the two scalar fields ν and λ.

However, there exist a second way to get the equations for ν and λ. We know that the Einstein field equations, $G^{ij}=0$, can be derived by varying the metric tensor in $\mathcal{L}_E := g^{1/2} R$, where $g := |\det g_{ij}|$. So if we build $\mathcal{L}_E$ from the line element given in Eq.2.2, and then vary the scalar fields ν and λ in the resulting Lagrangian, we obtain Euler-Lagrange equations equivalent to those obtained from $G^{ij} = 0$.

This observation concerning the Schwarzschild equations suggests an approach to deriving field equations from a Lagrangian of the form presented in Eq.2.1. We should begin by guessing a form of f suitable to the problem at hand. Plug that ansatz f into $\mathcal{L}$ along with φ, and then vary the fields in $\mathcal{L}$, just as we did in $\mathcal{L}_E$ above. The Lagrangian that we get in this manner will be called a hidden scalar Lagrangian. It will be a concomitant of φ, its derivatives, and perhaps the local coordinates of M, if f locally depends upon the $\chi^i$ coordinates as well as the $y_i$, as it will for those f's that generate FLRW metrics with $k \neq 0$.

This approach to constructing field equations for scalar-scalar field theories using the Horndeski Lagrangians [11], [12] (see also Deffayet, *et al.,* [13] and Kobayashi, *et al.,* [14] for these Lagrangians) was studied at great length in [9]. In the course of studying the hidden Lagrangians associated with the Horndeski Lagrangians I discovered a "true" scalar-scalar theory Lagrangian, $L_f$ which was related to the quartic and quintic Horndeski Lagrangians (*see,* Eqs.3.78, 3.82 and 3.86 in [9]). By $L_f$ being a "true" scalar-scalar Lagrangian I mean that it does not arise as the hidden Lagrangian associated with some scalar-tensor Lagrangian. $L_f$ is defined as follows. If f is the Cofinsler function of a scalar-scalar theory with scalar field φ, we let f* be the scalar field on M defined by f*:= f ∘ dφ ≡ dφ*f, and set

$$L_f := -\tfrac{1}{8}\kappa g^{\frac{1}{2}} \varphi f^{*-4} g^{ab} f^{*}_{,a} f^{*}_{,b}, \qquad \text{Eq.2.3}$$

where $\kappa \sim \ell^{-1}$ is a constant. We shall investigate the Lagrangian $L_f$ when f is chosen to be the flat Cofinsler function presented in Eq.1.7. Using this f in Eq.2.3 we easily find

$$L_f = \tfrac{1}{2}\kappa \varphi (\varphi')^{-3} (\varphi'')^2, \qquad \text{Eq.2.4}$$

assuming φ'>0, so that $g^{\frac{1}{2}} = |\det g_{ij}|^{\frac{1}{2}} = (\varphi')^3$.

At this point most savvy readers have probably recognized that $L_f$ is a non-degenerate, second-order Lagrangian, and hence possesses Ostrogradsky [15] type singularities (*cf.,* Woodard [16]). Thus many readers probably suspect that $L_f$ should be dropped from further consideration. But I say, let us not be too hasty. One of the problems associated with Lagrangians like $L_f$ *is* that they can lead to multiple vacuum states. This would be a good thing if we were looking for an equation that predicted the multiverse, since a multiverse might need a separate vacuum for each of the individual universes. So let's continue.

For a Lagrangian, L, which is second-order in φ = φ(t), the Ostrogradsky Hamiltonian is given by (*see,* Woodard [16])

$$H := P_1\varphi' + P_2\varphi'' - L, \qquad \text{Eq.2.5}$$

where

$$P_1 := \frac{\partial L}{\partial \varphi'} - \frac{d}{dt}\frac{\partial L}{\partial \varphi''}, \text{ and } P_2 := \frac{\partial L}{\partial \varphi''}. \qquad \text{Eq.2.6}$$

One useful property of the Ostrogradsky Hamiltonian is that it satisfies the following identity

$$\frac{dH}{dt} = -\varphi' \frac{\delta L}{\delta \varphi}. \qquad \text{Eq.2.7}$$

Thus, as is well-known, when the field equations are satisfied, H is a constant, and conversely, when H is constant, the field equations are satisfied. So the equation H= constant, is a first integral of our field equations. In particular we shall be interested in the solutions to H = 0, which I shall call the vacuum solutions to our sourceless scalar-scalar theory. This equation will be expressed in terms of $\varphi$ and its derivatives, and not in terms of canonical variables.

At this point I should remark that traditionally the vacuum solutions are taken to be those solutions to H = $H_0$, where $H_0$ is the smallest constant value that H can assume. In Appendix A of [9] I examine the Hamiltonian $H_f$ associated with $L_f$ and discover that $H_f$=constant, admits solutions for all choices of that constant. So there are no minimum energy solutions. Nevertheless, for the remainder of this paper I shall in general refer to the solutions of H = 0 as vacuum solutions to our source-less field equations.

Using Eqs.2.4-2.6 we find that the Hamiltonian $H_f$ associated with the Lagrangian $L_f$ is given by

$$H_f = -\kappa \frac{d}{dt}[\varphi(\varphi')^{-2}\varphi''].  \quad \text{Eq.2.8}$$

It is an elementary matter to show that all of the vacuum solutions to $H_f = 0$ are given by

$$\varphi = \alpha e^{\beta t} \quad \text{and} \quad \varphi = \gamma(\varepsilon_1 t + \varepsilon_2)^q  \quad \text{Eq.2.9}$$

where $\alpha \sim \ell^1$, $\beta \sim \ell^{-1}$, $\gamma \sim \ell^1$, $\varepsilon_1 \sim \ell^{-1}$, $\varepsilon_2 \sim \ell^0$ and $q \sim \ell^0$ are real numbers chosen so that $\varphi$ is well defined with $\varphi' > 0$. The exponential solutions provide us with examples of de Sitter geometries [17]. Of course, the constant $\varepsilon_1$ can be pulled out of the second expression for $\varphi$ and absorbed into $\gamma$. However, I shall not do that since it gives $\gamma$ peculiar units, effecting our later interpretation of $\varphi$ for collapsing universes in which $\varphi = -\gamma(\varepsilon_2 - \varepsilon_1 t)^q$. When q lies in the range $1 < q < 2$, we can use Eqs.1.8 and 2.9 to deduce that the metrics obtained using $\varphi = \gamma(\varepsilon_1 t)^q$, with $\varepsilon_1 > 0$ and $t \geq 0$, begin explosively at $t=0$; i.e., $\varphi'(0)=0$ and $\lim \varphi''(t) = \infty$. So these spaces begin expanding faster at $t=0$, than any exponential $t \to 0^+$ solution to $H_f = 0$ could. For the FLRW metric given in Eq.1.6 with $a(t) = \varphi'(t)$ we have

$$R = 6(\varphi')^{-2}[\varphi'\varphi''' + (\varphi'')^2 + k]  \quad \text{Eq.2.10a}$$

and

$$R^{abcd}R_{abcd} = 12\{(\varphi'''/\varphi')^2 + (\varphi')^{-4}[(\varphi'')^2 + k]^2\}  \quad \text{Eq.2.10b}$$

and thus the spaces with $\varphi=\gamma(\varepsilon_1 t)^q$ have a genuine curvature singularity at t=0, while this is not the case for the solutions generated by $\varphi = \alpha e^{\beta t}$, $t \in \mathbb{R}$.

Up to this point we have been dealing with the k=0, FLRW Cofinsler function given in Eq.1.7. A Cofinsler function that would generate the general FLRW metric given in Eq.1.5 with $a(t) = \varphi'(t)$ is given by

$$f(k) := -(y_t)^2 + (y_t)^{-2}((1-kr^2)(y_r)^2 + r^{-2}(y_\theta)^2 + r^{-2}\sin^{-2}\theta(y_\zeta)^2) . \qquad \text{Eq.2.11}$$

If we employ f(k) to build the Lagrangian presented in Eq.2.3 we obtain

$$L_{f(k)} = (1-kr^2)^{-\frac{1}{2}} r^2 \sin\theta \, L_f \qquad \text{Eq.2.12}$$

where $L_f$ is given by Eq.2.4. Combining Eqs.2.5, 2.6 and 2.12 shows us that the Hamiltonian, $H_{f(k)}$ generated by $L_{f(k)}$ is given by

$$H_{f(k)} = (1-kr^2)^{-\frac{1}{2}} r^2 \sin\theta \, H_f \qquad \text{Eq.2.13}$$

with $H_f$ as in Eq.2.8. So $H_{f(k)}$ and $H_f$ have the same vacuum solutions; *viz.,* those given in Eq.2.9. (They also have the same non-vacuum solutions (*see*, [9]).)

In the next section I shall discuss how some of the solutions to $H_{f(k)}=0$, presented in Eq.2.9, can be pieced together to obtain a multiverse, and how $L_{f(k)}$ can be used to to get us thinking about which universes might be more probable venues for matter.

**Section 3: Constructing Universes and Mutiverses**

In the standard description of inflationary models of the Universe (*see,* Guth [18], for an excellent discussion of these models, their origin and development prior to 1997), the Universe expands rapidly in an exponential manner for a short period of time from a state of thermodynamic equilibrium near t=0. After that burst there is a moderate, albeit somewhat accelerated, expansion for eternity. I shall demonstrate that we can obtain something like that using the solutions to $H_f$=0 given in Eq.2.9. I shall also show that we can get some solutions to stop expanding, and then shrink to as small a value as you wish for the scale factor φ'(t) at some time in the future.

All of the vacuum solutions of the Hamiltonian $H_f$ are presented in Eq.2.9. It is natural to at first restrict one's attention to the case where α>0, β>0, γ>0 and $\varepsilon_1$>0, since these give rise to expanding universes. However, the cases where α<0, β<0, γ<0 and $\varepsilon_1$<0 are also interesting, since they can represent collapsing universe solutions, with φ' still greater than 0. My contention is that we can combine these various vacuum solution to obtain universes in which the metric tensor is everywhere continuous, but experiences discontinuities in its derivatives as the universe jumps between different states of the scalar field φ, which can be construed as phase transitions of φ.

To begin with let us consider spaces for which $\varphi = \gamma(\varepsilon_1 t + \varepsilon_2)^q$. In these spaces φ' will experience a singularity at t=$-\varepsilon_2/\varepsilon_1$ when q<1, and will be flat when q=1. So we shall confine our attention to these types of solutions when q≥1. On the other hand solutions of the form $\varphi = \alpha e^{\beta t}$, are valid for all t∈ℝ, with φ' singularity free.

Consider the discontinuous, but piecewise $C^\infty$ solution to $H_f = 0$ (which is also a solution to $H_{f(k)} = 0$, for all k) given by

$$\varphi := \begin{array}{l} \varphi_1 = t, \text{ for } t<0, \\ \varphi_2 = \beta_1^{-1}\exp(\beta_1 t), 0 \leq t < t_1 \text{ and} \\ \varphi_3 = \alpha_2\exp(\beta_2 t), t_1 \leq t \end{array} \qquad \text{Eq.3.1}$$

where $t_1 > 0$ is chosen so that $\varphi_2'(t_1) = \varphi_3'(t_1)$, with $\beta_1 > 0$ and $\alpha_2 \beta_2 > 0$ (so that $\varphi' > 0$). This function $\varphi$ is clearly going to have discontinuities at $t=0$ and $t=t_1$, although $\varphi'$ will be continuous at these points if we use left and right hand derivatives. As a result we can use Eq.3.1 in conjunction with Eq.1.8 (for the flat FLRW space) or Eqs.1.4 and 2.11 (for the more general FLRW space) to construct a universe that is Minkowskian (when k=0) from $t=-\infty$ to $t=0^-$, begins to expand exponentially at $t=0$ with the scale factor $\varphi'(0)=1$, and then jumps at time $t=t_1$ to another exponentially expanding (or contracting if $\beta_2 < 0$) universe in such a way that the metric tensor remains continuous. In addition the vacuum field equations are satisfied at all times except when $t=0$, and $t=t_1$, where the metric tensor is not differentiable. This solution would be my version of the type of solution suitable for Guth's inflationary universe.

Note that due to Eq.2.10a we see that for the scalar field presented in Eq.3.1: (i) the scalar curvature jumps from $R=0$ to $R=12\beta_1^2$, as we pass from $\varphi_1$ to $\varphi_2$ at $t=0$; and (ii) R jumps from $12\beta_1^2$ to $12\beta_2^2$, as we pass from $\varphi_2$ to $\varphi_3$ at $t=t_1$. So if $\beta_2 = -\beta_1$, both R and $R_{abcd}R^{abcd}$ would be continuous across $t=t_1$. In Guth's inflationary models $\beta_1$ is quite large, so the jump in R as time passes through $t=0$ could be quite large, which would not be too surprising when the universe begins. But in Guth's models $\beta_2$ would be close to 0, so there would be a shocking reduction in the value of R as time passes through $t=t_1$, which, in the context of the present theory, should leave a mark on the universe.

If $\alpha_2$ and $\beta_2$ in Eq.3.1 are both positive, then we could then add a fourth branch to the function $\varphi$ given in Eq.3.1 of the form

$$\varphi := \varphi_4 = -\alpha_3 \exp(-\beta_3 t) , \; t_2 \leq t \qquad \text{Eq.3.2}$$

where $\varphi_3'(t_2) = \varphi_4'(t_2)$, with $\varphi_3$ now having the domain $t_1 \leq t < t_2$. When $\alpha_3 > 0$, $\beta_3 > 0$, this fourth branch of $\varphi$ would correspond to an exponentially decreasing branch of the universe. If we measure length in terms of the reduced Planck length L*, this branch could continue to decrease forever, which really would not make sense from a quantum gravity point of view. It would seem more reasonable for this branch to stop when $t=t_c$, the "cutoff" or "crunch time" of the universe, where $\varphi'(t_c) = 1$. After reaching that point the universe could just start all over again on a version of branch two given in Eq.3.1. This process could go on for eternity, and there is really no reason for it not to have gone on prior to branch two of the $\varphi$ solution given in Eq.3.1. So instead of beginning with $\varphi = t$ in Eq.3.1, we could begin with $\varphi$ being given by the tail end of a solution of the form given in Eq.3.2.

For later use let us set

$\mathbf{P_3} := \{(\beta_1,\alpha_2,\beta_2) \in \mathbb{R}^+ \times \mathbb{R}^2 \mid \exists\, t_1 > 0,\text{ with } \exp(\beta_1 t_1) = \alpha_2\beta_2\exp(\beta_2 t_1)\}$  Eq.3.3

$\mathbf{P_5} := \{(\beta_1,\alpha_2,\beta_2,\alpha_3,\beta_3) \in \mathbb{R}^+ \times \mathbb{R}^4 \mid \exists\, t_c > t_2 > t_1 > 0,\text{ with } \exp(\beta_1 t_1) = \alpha_2\beta_2\exp(\beta_2 t_1),$

$\alpha_2\beta_2\exp(\beta_2 t_2) = \alpha_3\beta_3\exp(\beta_3 t_2),\text{ and } \alpha_3\beta_3\exp(\beta_3 t_c) = 1,\text{ with } \beta_i \neq \beta_{i+1},\, i=1,2\}$.  Eq.3.4

Strangely enough functions with discontinuities, but "continuous first derivatives," will arise frequently when we try to piece together vacuum solutions of $H_{f(k)}$, and I shall refer to them as functions of class $C^{k,1}$. To be precise I define such functions as follows. Let $\psi: \mathbb{R} \to \mathbb{R}$ be defined on an interval of the form $(a,b) \subset \mathbb{R}$ (where a could be $-\infty$ and b could be $+\infty$), and let $\{P_i\}_{i \in I}$ be distinct points in (a,b), where $I \subset \mathbb{Z}$ (the set of integers) is a non-empty indexing set, which could be countably infinite. I shall assume that the set $\{P_i\}_{i \in I}$ does not have a limit point, and thus $D := (a,b) \setminus \{P_i\}_{i \in I}$ is an open subset of $\mathbb{R}$. $\psi$ is said to be of class $C^{k,1}$ on (a,b) ($k \in \mathbb{N}$, the set of positive integers, or $k = \infty$) if:

(i) $\psi$ is discontinuous at the points $\{P_i\}_{i \in I}$,

(ii) $\psi$ is of class $C^k$ on D, and

(iii) $\forall\, P_i$, with $i \in I$, $\lim_{t \to P_i^-} \psi'(t) = \lim_{t \to P_i^+} \psi'(t) =: \psi'(P_i)$.

I shall now present examples of some universes that begin catastrophically. If $1 < q < 2$ then the solution $\varphi = \gamma(\varepsilon_1 t)^q$, $t \geq 0$, begins with $\varphi'(0) = 0$, and $\varphi''(0) = \infty$. This solution begins explosively at $t = 0$, with a vertical tangent vector for the graph of $\varphi'$ versus t. As time increases the graph of $\varphi'$ passes through the graphs of $\varphi_e'$ for a myriad of solutions to $H_f=0$ for which $\varphi_e = \alpha\exp(\beta t)$, with $\varphi_e'=\alpha\beta\exp(\beta t)$, where $\alpha>0$, and $\beta>0$.

My contention is that when our original universe, with $\varphi' = q\gamma\varepsilon_1(\varepsilon_1 t)^{q-1}$, crosses a universe with $\varphi_e' = \alpha\beta\exp(\beta t)$ at some time $t_1$, it can "jump" from the original $\varphi$ state, to the exponential $\varphi_e$ state, when $q\gamma\varepsilon_1(\varepsilon_1 t_1)^{q-1} = \alpha\beta\exp(\beta t_1)$. Due to Eq.1.8, this condition guarantees that the metric of the universe is continuous across the jump. However, the first and second derivatives of the metric, in particular the scalar curvature R (see Eq.2.10a), is in general not continuous across the jump at time $t = t_1$.

Let's assume that $\alpha$, $\beta$, $\gamma$ and $\varepsilon$ have been chosen so that when $1<q<2$, the equation $q\gamma\varepsilon(\varepsilon t)^{q-1} = \alpha\beta\exp(\beta t)$ has two distinct solutions $t_1$ and $t_2$ with $0<t_1<t_2$. If $\varphi$ is chosen so that

$$\varphi := \begin{cases} \varphi_1 = \gamma(\varepsilon t)^q, & 0 \leq t < t_1, \\ \varphi_2 = \alpha\exp(\beta t), & t_1 \leq t < t_2, \\ \varphi_3 = -\gamma(\varepsilon(2t_2-t))^q, & t_2 \leq t \leq 2t_2, \end{cases} \quad \text{Eq.3.5}$$

then due to Eq.1.8, $\varphi$ gives us a continuous metric tensor for $0<t<2t_2$, which satisfies the field equation $H_f=0$ (as well as $H_{f(k)}=0$) for $0<t<2t_2$, except at $t= t_1, t_2$. The solution given in Eq.3.5 is of class $C^{\infty,1}$ on $(0, 2t_2)$, and has genuine curvature singularities at $t=0$ and $t=2t_2$ where due to Eq.2.10a

$$\lim_{t\to 0^+} R = \lim_{t\to 2t_2^-} R = \infty.$$

Thus the geometry generated on $[0, 2t_2]$ by this solution begins with a curvature singularity at $t=0$ and also ends with one at $t=2t_2$.

I shall admit that some people may regard it as strange for me to be seriously considering these "mix and match" universes, made from vacuum solutions in which φ is discontinuous at various instants of time. But I believe it to be justifiable, so long as the metric tensor is continuous. As mentioned previously, these jumps in φ can be construed as phase transitions, and we have no reason to believe that they do not exist. The discontinuities in R and $R^{abcd}R_{abcd}$ which can accompany the discontinuities in φ might leave an impression upon matter in the universe and could be quite profound, especially to life, if they are at all large. Such discontinuities could act like a "hammer blow" to the universe.

Now the natural question to ask is: When t>0 how does the initial exploding state of the solution given in Eq.3.5 pick which exponential state it wants to jump into when the φ''s are equal? A similar question applies to the exponential solutions obtained using Eq.3.1 and 3.2. Well, one way to avoid this problem of making choices is to assume that everything that can happen does. So we are dealing with multiple universes; *i.e.*, a multiverse. I shall make this thought a bit more precise.

Suppose that ∀ choice of non-negative α, β, γ, ε∈ℝ, and q with 1<q<2, we define

$$\varphi_{\alpha,\beta} := \alpha\exp(\beta t), \text{ and } \varphi_{\gamma,\varepsilon,q} := \gamma(\varepsilon t)^q, \qquad \text{Eq.3.6}$$

where $\varphi'_{\alpha,\beta} > 0$, and $\varphi'_{\gamma,\varepsilon,q} > 0$. Let

$$\mathbf{P_4((1,2))} :=$$

$$\{\,(\alpha,\beta,\gamma,\varepsilon,q)\in(\mathbb{R}^+)^4 \times (1,2)\mid \exists\, t_2>t_1>0,\text{ with } \varphi'_{\alpha,\beta}(t_i)=\varphi'_{\gamma,\varepsilon,q}(t_i), i=1,2\}. \qquad \text{Eq.3.7}$$

Due to the nature of the exponential function we know that $\mathbf{P_4((1,2))}$ is non-empty. $\mathbf{P_4((1,2))}$, $\mathbf{P_3}$ and $\mathbf{P_5}$ (*see*, Eqs.3.3 and 3.4) will be the space of parameters labelling the universes in the multiverse which I shall now construct. To that end I need to construct a 1-dimensional manifold which represents a notion of time that multifurcates when t=0.

Back in the early 1970's I encountered a very interesting non-Hausdorff, 1-dimensional manifold, which I thought might prove useful to me some day. This manifold occurred as an exercise in Brickell & Clark [19] (*see,* problem 3.2.1 on page 40 of [19]). I shall describe a slight modification of this manifold, and then show how this manifold can be generalized to build multiverses consisting of many of the universes generated as solutions to $H_f=0$ and $H_{f(k)}=0$.

Consider the subset $T$ of $\mathbb{R}^2$ defined by $T := \underset{\zeta\in\mathbb{R}}{\cup}\, U_\zeta$, where

$U_\zeta := \{\,(t,0)\mid t<0\,\} \cup \{\,(t,\zeta)\mid t\geq 0\,\}$.

$\forall\,\zeta\in\mathbb{R}$ I define a 1-dimensional chart $t_\zeta: T\to\mathbb{R}$ with domain $U_\zeta$ by: $t_\zeta((t,0)) := t$, and $t_\zeta((t,\zeta)) := t$. If $\zeta, \eta\in\mathbb{R}$, then $t_\zeta$ and $t_\eta$ are clearly $C^\infty$ related. The collection of charts $C(T) := \{t_\zeta\}_{\zeta\in\mathbb{R}}$ defines a $C^\infty$ atlas of $T$ into $\mathbb{R}$, and hence $T$ is a differentiable manifold of dimension 1 with its $C^\infty$ structure determined by $C(T)$. ( I am using Brickell & Clark's [19] definition of a differentiable manifold here, which does not require $T$ to be endowed with a topological structure before the charts are defined.) The domains of the charts of the complete atlas of $T$ generated by $C(T)$, determines a basis for a topology on $T$ which we shall take as the manifold topology. In terms of this topology $T$ is connected and satisfies the $T_1$ separation axiom. However, if $\zeta \neq \eta$ then it is impossible to separate the points $(0,\zeta)$ and $(0,\eta)$ by disjoint open sets. Thus the manifold topology of $T$ is non-Hausdorff. Since $\mathbb{R}$ is uncountable, the atlas $C(T)$ is also uncountable. Due to Proposition 3.3.3 in [19], this implies that $T$ does not admit a countable basis for its topology. I shall refer to the coordinate domains $U_\zeta$ as branches of the manifold $T$.

I define a function t: $T \to \mathbb{R}$ by stipulating that t:= $t_\zeta$ on the domain of each chart $U_\zeta$. t is evidently well-defined and differentiable. We can think of t as the global time function on $T$, which multifurcates at time t = 0, since $t^{-1}(0)$ is $\{ (0,\zeta) \mid \zeta \in \mathbb{R}\}$, while if $\gamma<0$, then $t^{-1}(\gamma) = (\gamma,0)$.

Before I develop the time manifold we require for scalar-scalar field theories I would like to show how $T$ can be modified to permit time travel between different branches of $T$. To help you envision what I am doing, recall the telephone switch board operators of yesteryear. In movies from the 1930's you would often see a phone operator sitting in front of a vertical switch board, and connect calls from one party to another using a wire extending from the caller's port in the switch board, and going to the the recipient's plug in that board. I shall do something similar with $T$.

Let $(t,\zeta)$, and $(t,\zeta,\eta)$ denote the standard coordinates of $\mathbb{R}^2$ and $\mathbb{R}^3$, respectively. We embed $T \subset \mathbb{R}^2$ into $\mathbb{R}^3$ as the $\eta=0$ plane. Consider the two points $(t_1,\zeta_1,0)$ and $(t_2,\zeta_2,0) \in T$, where $0 < t_1 < t_2$, and let

$$U(t_1,\zeta_1;t_2,\zeta_2) := \{(t,\zeta,0) \in U_{\zeta 1} | t < t_1\} \cup \{\text{line segment in } \eta=1 \text{plane from } (t_1,\zeta_1,1) \text{ to } (t_2,\zeta_2,1)\}$$
$$\cup \{(t,\zeta,0) \in U_{\zeta 2} | t > t_2\} \ .$$

Set $T(t_1,\zeta_1;t_2,\zeta_2) := T \cup U(t_1,\zeta_1;t_2,\zeta_2)$ and let $t(t_1,\zeta_1;t_2,\zeta_2)$ be the restriction of the projection $(a,b,c) \to a$ to $U(t_1,\zeta_1;t_2,\zeta_2)$. Clearly $t(t_1,\zeta_1;t_2,\zeta_2)$ is a chart for $T(t_1,\zeta_1;t_2,\zeta_2)$ and is $C^\infty$ related to the charts of $C(T)$. Hence $C(T) \cup \{t(t_1,\zeta_1;t_2,\zeta_2)\}$ defines a $C^\infty$ structure on $T(t_1,\zeta_1;t_2,\zeta_2)$ of dimension 1. From our construction it is evident that one can travel forward in time from the coordinate domain $U_{\zeta 1}$ to $U_{\zeta 2}$, leaving from the point $(t_1,\zeta_1)^-$ and arriving at the point $(t_2,\zeta_2)^+$ some time ($=t_2-t_1$) later. It should also be clear how one could travel in time between other points in $T$.

Next I shall describe a generalization of the 1-dimensional manifold $T$ that will be required for our multiverse. This manifold will also be 1-dimensional, and will embody multifurcating time.

Let $P_n$ be a subset of $\mathbb{R}^n$. In our applications $P_n$ will comprise the parameters that characterize different vacuum solutions, as in Eqs.3.3, 3.4 and 3.7. $T(P_n)$ will denote the subset of $\mathbb{R}^{n+1}$ defined by $T(P_n) := \bigcup_{p \in P_n} U_p(P_n)$, where if $p = (p_1, \ldots, p_n) \in P_n$ then

$$U_p(P_n) := \{ (t,0,\ldots,0) \in \mathbb{R}^{n+1} \mid t<0 \} \cup \{ (t, p_1, \ldots, p_n) \mid t \geq 0 \}.$$

$\forall p \in P_n$, I define a 1-dimensional chart $t_p$ with domain $U_p(P_n)$ by

$$t_p((t,0,\ldots,0)) := t \text{ and } t_p((t, p_1, \ldots, p_n)) := t.$$

$U_p(P_n)$ will be called a branch of $T(P_n)$. Due to my previous remarks it should be evident that the collection of charts $C(T(P_n)) := \{t_p \mid p \in P_n\}$ determines a $C^\infty$ structure of dimension 1 on $T(P_n)$. With this structure $T(P_n)$ is a differentiable manifold of dimension 1. In terms of the manifold topology $T(P_n)$ is connected and satisfies the $T_1$ separation axiom, but is non-Hausdorff. This manifold admits a countable basis for its topology if and only if $P_n$ is countable. I define the global time function t on $T(P_n)$ in a manner analogous to how it was defined on $T$. Once again we see that time essentially fractures as we pass through $t=0$, but even so the manifold remains connected. It should be evident how $T(P_n)$ could be modified to allow time travel between two points $(t_1,p_1)$, $(t_2,p_2) \in T(P_n)$, when $0 < t_1 < t_2$.

The spaces $T(P_n)$ can be used to construct various mulitverses in the following manner. If we choose k=0, then I shall denote the underlying manifold of three multiverses of interest to us by

$$\mathbf{MV_{k=0}(P_3)} := T(P_3) \times \mathbb{R}^3 \qquad \text{Eq.3.8}$$

$$\mathbf{MV_{k=0}(P_5)} := T(P_5) \times \mathbb{R}^3 \qquad \text{Eq.3.9}$$

and
$$\mathbf{MV_{k=0}(P_4((1,2)))} := T(P_4((1,2))) \times \mathbb{R}^3 \,. \qquad \text{Eq.3.10}$$

Geometric structures consistent with our solutions to $H_f=0$ can be defined on each of these spaces as follows.

Let $p:=(\beta_1,\alpha_2,\beta_2) \in \mathbf{P_3}$ as defined in Eq.3.3, and let $t_p$ be the corresponding chart of $T(\mathbf{P_3})$ with domain $U_p(\mathbf{P_3})$. If t denotes the global time function of $T(\mathbf{P_3})$, and (u,v,w) denotes the standard chart of $\mathbb{R}^3$ then on the branch $U_p(\mathbf{P_3}) \times \mathbb{R}^3$ of $\mathbf{MV_{k=0}(P_3)}$

$$\varphi := \begin{cases} t, & \text{if } t<0 \\ \beta_1^{-1} \exp(\beta_1 t), & \text{if } 0 \leq t < t_1 \\ \alpha_2 \exp(\beta_2 t), & \text{if } t_1 \leq t \,. \end{cases} \qquad \text{Eq.3.11}$$

$\varphi$ is evidently well defined on $\mathbf{MV_{k=0}(P_3)}$, since if p, p'$\in \mathbf{P_3}$ then on the intersection of the branches $U_p(\mathbf{P_3}) \times \mathbb{R}^3$ with $U_{p'}(\mathbf{P_3}) \times \mathbb{R}^3$, $\varphi=t$. $\varphi=\varphi(t)$ is of class $C^{\infty,1}$ on each $U_p(\mathbf{P_3})$, and satisfies the equation $H_f=0$, except at the points where it is discontinuous. If we combine the value of $\varphi$ given in Eq.3.11 with the formula for $g^{ij}$ given in Eq.1.8, we obtain a well-defined continuous metric tensor on $\mathbf{MV_{k=0}(P_3)}$, which is of class $C^{\infty}$ except where $\varphi$ is discontinuous.

Similarly, if $p=(\beta_1,\alpha_2,\beta_2,\alpha_3,\beta_3)$ is any point in $\mathbf{P_5}$ (*see*, Eq.3.4), then on $U_p(\mathbf{P_5}) \times \mathbb{R}^3$

$$\varphi := \begin{cases} t, & \text{if } t<0, \\ \beta_1^{-1} \exp(\beta_1 t), & \text{if } 0 \leq t < t_1 \\ \alpha_2 \exp(\beta_2 t), & \text{if } t_1 \leq t < t_2 \\ \alpha_3 \exp(\beta_3 t), & \text{if } t_2 \leq t < t_c, \\ t, & \text{if } t_c \leq t \,. \end{cases} \qquad \text{Eq.3.12}$$

Just as above, $\varphi$ defined by Eq.3.12 is well-defined on all of $\mathbf{MV_{k=0}(P_5)}$ and satisfies

$H_f$=0, except where φ is discontinuous. φ also generates a continuous metric tensor on all of $\mathbf{MV}_{k=0}(\mathbf{P_5})$, which is $C^\infty$, except where φ experiences discontinuities.

Lastly, in keeping with the above work, if p=(α,β,γ,ε,q) is an arbitrary point in $\mathbf{P_4((1,2))}$ (*see*, Eqs.3.6 and 3.7), then on the branch $U_p(\mathbf{P_4((1,2))})\times\mathbb{R}^3$

$$\varphi := \begin{cases} 0, & \text{if } t<0, \\ \gamma(\varepsilon t)^q, & \text{if } 0\leq t<t_1 \\ \alpha\exp(\beta t), & \text{if } t_1\leq t<t_2, \\ -\gamma(\varepsilon(2t_2-t))^q, & \text{if } t_2\leq t\leq 2t_2. \end{cases} \qquad \text{Eq.3.13}$$

These expressions for φ on the coordinate domains branches of $\mathbf{MV}_{k=0}(\mathbf{P_4((1,2))})$, evidently piece together to give a well-defined scalar field on $\mathbf{MV}_{k=0}(\mathbf{P_4((1,2))})$, which is of class $C^{\infty,1}$ and satisfies $H_f$=0, when $0<t<2t_2$, except at $t=t_1$ and $t=t_2$. Unlike the two cases examined above, when t<0, Eqs.1.8 and 3.13 tell us that the "metric" on $\mathbf{MV}_{k=0}(\mathbf{P_4((1,2))})$, is $ds^2 = -dt^2$, which is degenerate. However, when t>0 we obtain a Lorentzian metric which is continuous, and piecewise of class $C^\infty$ on $0<t<2t_2$. As mentioned previously, this metric has curvature singularities as $t\to 0^+$ and $t\to 2t_2^-$.

In an analogous manner we can use the solutions to $H_{f(k)}$=0, to define scalar fields and geometries on the multiverses

$$\mathbf{MV}_{k>0}(\mathbf{P_3}) := T(\mathbf{P_3})\times S_k,\ \mathbf{MV}_{k>0}(\mathbf{P_5}) := T(\mathbf{P_5})\times S_k,$$
and
$$\mathbf{MV}_{k>0}(\mathbf{P_4((1,2))}) := T(\mathbf{P_4((1,2))})\times S_k \qquad \text{Eq.3.14}$$

where $S_k$ is a 3-sphere of radius $R_0 = k^{-1/2}$. Similar definitions apply to the multiverses generated by the k<0 solutions to $H_{f(k)}$=0.

It should be noted that in each of the multiverses I defined above gravity really does not exist until t>0. In particular the spaces $MV_0(P_3)$ and $MV_0(P_5)$ are just Minkowski space when t<0, in which standard quantum field theory can be used to describe the behavior of matter. While for $MV_k(P_4)$ (k arbitrary) when t<0 the physical distance between any two points in a t=constant slice is zero. So it is not really clear what matter would be doing in $MV_k(P_4)$ when t<0, if matter were actually present.

The multiverses presented above are quite different than those introduced by Linde in [20], and discussed in Guth [18]. The construction of those multiverses combined notions from elementary particle physics, along with a scalar field (the inflaton field), and Einstein's equations with cosmological constant. I have presented a more mathematical approach to the construction of multiverses. One would think that since about 95% of the energy in our universe comes from dark energy and dark matter, it should be possible to construct multiverses by just considering those sources, with no thought about ordinary matter. That is what I have tried to do here.

The multiverses I have introduced might give us some insight into why our universe is dominated by matter, and does not contain equal amounts of matter and antimatter without appealing to the physics of elementary particles. Admittedly, the multiverses I have described are built from sourceless solutions of the field equations, but let us suppose that when matter is present we can still construct a multiverse of the form $\mathbf{MV}(P_n)$ for some parameter space $P_n$ associated with the solutions, and time mutifurcating at t=0.  When t<0 all of the particles in the  multiverse  will be  in a single $\mathbb{R}^3$ or $S_k$.  Let's  suppose that this infinite sea of particles is comprised of "equal" amounts of matter and antimatter.  Then when t=0, and time multifurcates, each quantum of matter is forced to go into some slice of the multiverse. We can only speculate on how particles make the decision where to go, and this topic will be discussed momentarily.  But in any case, it would seem highly unlikely that after t=0 all of the particles in the various slices of $\mathbf{MV}(P_n)$ are equally balanced between matter and antimatter. So during the period of time near $t=0^+$ one would expect that there would be a lot of annihilation between matter and antimatter, leaving a preponderance of one or the other as time evolves in each branch of $\mathbf{MV}(P_n)$.  One problem with this scenario is that  when t<0, the original sea of equal amounts of matter and antimatter was electrically neutral.   Then when these particles start spilling into the slices of $\mathbf{MV}(P_n)$ the resulting mix would more than likely have a net electrical charge after all the matter and antimatter that can annihilate near $t=0^+$ does so.  Our universe appears to violate this scenario since it does not seem to possess a net charge, although that net charge could be small, spread throughout the Universe, and thus be unnoticeable.

    I briefly mentioned above the problem of how particles moving through the

moment t=0 decide which branch of a multiverse it should enter. This leads us to wonder if it is possible to determine the distribution of particles as it pours into to the various branches of φ; will some branches be more likey to get more particles than other branches? The natural quantity to consider in this regard is the action of φ associated with the various branches. For recall that the solutions to the field equation for φ are intended to be extremals of this action.

With that in mind, suppose that φ is a solution of $H_{f(k)}=0$. I define the action associated with φ by

$$I_{f(k)}(\varphi) := \int L_{f(k)}(\varphi) d^4x \qquad \text{Eq.3.15}$$

where the integral is taken over a 4-dimensional region in a branch of the scalar field φ. Since we are dealing with FLRW Cofinsler functions we know that $L_{f(k)}(\varphi)$ is given by Eqs.2.4 and 2.12. Thus the integral appearing in Eq.3.15 can be broken into an integral over a compact region in a t=constant slice (whose value, $V_k \sim \ell^3$, does not depend upon the slice we choose), and an integral over an interval of time. *E.g.,* when k>0, the region can be taken to be the entire t = constant slice, for which $V_k=2\pi^2 R_0^3$, where $k=R_0^{-2}$. So in general

$$I_{f(k)}(\varphi) = V_k \int L_f(\varphi(t)) dt \qquad \text{Eq.3.16}$$

where the time integral is taken over an interval in the domain of φ. We shall examine the value of this integral for various φ solutions.

To begin, let us consider the scalar field defined for all time by Eq.3.11. Using Eq.2.4 we find that

$$L_f(\varphi=\alpha e^{\beta t}) = \tfrac{1}{2}\kappa\beta, \text{ and } L_f(\varphi=\gamma(\varepsilon_1 t+\varepsilon_2)^q) = \tfrac{1}{2}\kappa\varepsilon_1(q-1)^2[q(\varepsilon_1 t+\varepsilon_2)]^{-1}. \qquad \text{Eq.3.17}$$

When the expression for $\varphi$ given in Eq.3.11 is used in Eq.3.16 we discover that $I_f(\varphi) = \infty$, which is usually looked upon unfavorably as an extreme value for the action integral. However, we obtained this result only because we permitted the time integral in Eq.3.16 to extend over the entire domain of $\varphi$. Had we chosen a compact interval of time, the integral would be finite. The solution given in Eq.3.11 is the one most compatible with Guth's inflationary model, where the expansion of the Universe continues forever, and the value of the action grows right along with it.

The next solution of interest is the one given in Eq.3.12, which is well defined on any branch of the multiverse $\mathbf{MV_k(P_5)}$, not just for k=0. In this solution the universe has $\varphi'=1$ until t=0, after which it expands exponentially (rapidly if $\beta_1 \gg 0$), then coasts exponentially (more slowly if $\beta_2 \ll \beta_1$), and then collapses back down to $\varphi'=1$ at $t=t_c$. Using Eqs.3.16 and 3.17 we easily find that

$$I_{f(k)}(\varphi) = \tfrac{1}{2}\kappa V_k(\beta_1 t_1 + \beta_2(t_2-t_1) + \beta_3(t_c-t_2))$$

and hence

$$I_{f(k)}(\varphi) = \tfrac{1}{2}\kappa V_k((\beta_1-\beta_2)t_1 + (\beta_2-\beta_3)t_2 + \beta_3 t_c). \qquad \text{Eq.3.18}$$

Since $\varphi$ is of class $C^{\infty,1}$ we can use Eq.3.12, evaluated at times $t_1$, $t_2$ and $t_c$, to obtain

$\exp(\beta_1 t_1) = \alpha_2 \beta_2 \exp(\beta_2 t_1)$, $\alpha_2 \beta_2 \exp(\beta_2 t_2) = \alpha_3 \beta_3 \exp(\beta_3 t_2)$ and $\alpha_3 \beta_3 \exp(\beta_3 t_c) = 1$.

Upon taking the ln of each of these three equations we find that

$(\beta_1 - \beta_2)t_1 = \ln(\alpha_2 \beta_2)$, $(\beta_2 - \beta_3)t_2 = \ln(\alpha_3 \beta_3/(\alpha_2 \beta_2))$ and $\beta_3 t_c = -\ln(\alpha_3 \beta_3)$.

When these equations are combined with Eq.3.18 we discover that $I_{f(k)}(\varphi) = 0$, which seems to be an appropriate value for an extremum of the action integral.

So for this class of exponential solutions with three segments, two expanding and one contracting, each solution yields the value 0 for $I_{f(k)}(\varphi)$. Thus at first we are led to suspect that each branch of the multiverse **$MV_k(P_5)$** should be allocated about the same number of particles at t=0.

The last solution I wish to consider comes from the multiverse **$MV_k(P_4(1,2))$**. Let $\varphi$ be the solution given in Eq.3.5, with $0 \leq t \leq 2t_2$, and $(\alpha,\beta,\gamma,\varepsilon,q) \in \mathbf{P_4((1,2))}$. To use Eqs.3.16 and 3.17 to compute the action integral we need to get around the singularities at t=0 and t=$2t_2$. To that end I suggest we evaluate the time integral from t=$\mu \in (0, t_2)$ to t=$2t_2 - \mu$, and then take the limit as $\mu \to 0^+$. Upon doing this we find that the integrals over the first and last parts of $\varphi$ cancel leaving us with

$$I_{f(k)}(\varphi) = \kappa V_k \beta(t_2 - t_1) = \kappa V_k (q-1)\ln(t_2/t_1) \geq 0, \qquad \text{Eq.3.19}$$

where $V_k$ is the volume of a fixed compact region in a t=constant slice of the universe. Thus we see that if the exponential part of $\varphi$ were removed, so that $t_1 = t_2$, then $I_{f(k)}(\varphi) = 0$. When this is done the universe we get is analogous to particles popping into and out of existence in quantum field theory. These "bump" universes in $\mathbf{MV_k(P_4(1,2))}$ are the only ones which yield $I_{f(k)}(\varphi)=0$. However, we see from Eq.3.19 that it should be possible to find universes in $\mathbf{MV_k(P_4(1,2))}$ in which the middle exponential branch exists and for which $I_{f(k)}(\varphi)$ is as close to zero as we wish.

The fact that we can construct universes with $I_{f(k)}(\varphi)=0$ seems a bit disconcerting. For example, we can take an exponential solution $\varphi=\alpha e^{\beta t}$, $t_i \leq t \leq t_f$ and adjoin to it an exponential "bump," given by

$$\varphi_b := \begin{array}{ll} \varphi = \alpha\exp(\beta t), & t_i \leq t < t_1 \\ \varphi_1 = \alpha_1\exp(\beta_1 t), & t_1 \leq t < t_2 \\ \varphi_2 = \alpha_2\exp(\beta_2 t), & t_2 \leq t < t_3 \\ \varphi = \alpha\exp(\beta t), & t_3 \leq t \leq t_f \end{array} \qquad \text{Eq.3.20}$$

where $t_i<t_1<t_2<t_3<t_f$, and $\varphi'(t_1)=\varphi_1'(t_1)$, $\varphi_1'(t_2)=\varphi_2'(t_2)$, $\varphi_2'(t_3)=\varphi'(t_3)$, with all $\alpha$'s and $\beta$'s chosen so that $\varphi_b' >0$ where it is continuous. It can be shown using an argument similar to the one employed to compute $I_{f(k)}$ for the $\varphi$ solution given in Eq.3.12, that the value of the action $I_{f(k)}$ will be the same for $\varphi$ and $\varphi_b$ when computed for $t_i \leq t \leq t_f$. Similarly we could add an exponential "dip" to $\varphi$ without effecting $I_f$. Thus any number of bumps and dips could be added to the exponential portion of any universe in our multiverses without effecting the action for that universe. Now we do not see our universe expanding and contracting at random all of the time. So we might surmise that multiverses which permitted the existence of bumps and dips which did not change the value of the action, would be physically unrealistic, if we regard our universe as being in some sense typical. The problem here might be our original choice of Lagrangian. For when one does the calculations for these bumps and dips, the reason everything seems to cancel out just right is because the value of the action for a solution of the form $\varphi=\alpha e^{\beta t}$ is linear in $\beta$. There are several things which we can do to modify $L_f$ to change this situation.

Consider the following two Lagrangians

$$L_{1,f} := -\kappa_1 (g)^{\frac{1}{2}} \varphi^a |f^*|^b g^{ij}\varphi_{,i}\varphi_{,j} \quad \text{and} \quad L_{2,f} := \kappa_2 (g)^{\frac{1}{2}} \varphi^r |f^*|^s g^{ij} f^*_{,i}\varphi_{,j} \qquad \text{Eq.3.21}$$

where $\kappa_1$ and $\kappa_2$ are positive dimensioned constants. a,b,r,s are real numbers to be determined so that the Euler-Lagrange equations associated with $L_{1,f}$ and $L_{2,f}$ have $\varphi=\alpha e^{\beta t}$ as their solution with $L_{1,f}(\varphi=\alpha e^{\beta t}) \propto \beta^2$ and $L_{2,f}(\varphi=\alpha e^{\beta t}) \propto \beta^2$. If we evaluate $L_{1,f}$ for the flat FLRW metric given in Eq.1.8 we find that

$$L_{1,f} = \kappa_1 \varphi^a (\varphi')^{2b+5} . \qquad \text{Eq.3.22}$$

We want $L_{1,f}$ to be proportional to $\beta^2$ when evaluated for $\varphi=\alpha e^{\beta t}$. Using Eq.3.22 we see that this will be the case provided we choose $a=-2$ and $b=-3/2$, and so $L_{1,f}$ becomes

$$L_{1,f} = -\kappa_1 g^{\frac{1}{2}} \varphi^{-2} |f^*|^{-3/2} g^{ij}\varphi_{,i}\varphi_{,j} \qquad \text{Eq.3.23}$$

with $\kappa_1 \sim \ell^0$. In addition it is a simple matter to demonstrate that when f is given by Eq.1.7 the general solution of $\frac{\delta L_{1,f}}{\delta \varphi} = 0$, is given by $\varphi=\alpha e^{\beta t}$, where $\alpha,\beta \in \mathbb{R}\setminus 0$. The same result applies for the Cofinsler function f(k) given in Eq.2.11.

When a similar analysis is applied to the Lagrangian $L_{2,f}$ we find that when r and s are chosen to achieve the desired results, $L_{1,f}$ and $L_{2,f}$ agree up to a constant multiple. Hence we only need to consider $L_{1,f}$ in what follows.

At this juncture we need to see how $L_{1,f}$ affects the action when evaluated for bumps and dips in the evolution of an exponential portion of the universe. To that end let's consider the two scalar fields ( φ upper, denote $\varphi_U$ and φ lower, $\varphi_L$ ) given by

$$\varphi_U := \begin{cases} \varphi_1 = \alpha_1 \exp(\beta_1 t), & t_1 \leq t < t_2, \\ \varphi_2 = \alpha_2 \exp(\beta_2 t), & t_2 \leq t \leq t_3, \end{cases}$$

$$\varphi_L := \varphi_3 = \alpha \exp(\beta t), \quad t_1 \leq t \leq t_3,$$

where $\varphi_1'(t_1) = \varphi_3'(t_1)$, $\varphi_1'(t_2) = \varphi_2'(t_2)$ and $\varphi_2'(t_3) = \varphi_3'(t_3)$. We shall assume that $\beta_1 > \beta$ and $\beta > \beta_2$, which guarantees that the graph of $\varphi_U'$ lies above the graph of $\varphi_L'$, and that these graphs meet twice. It is important to note that by construction

$$\alpha_1 \beta_1 < \alpha \beta < \alpha_2 \beta_2, \qquad \text{Eq.3.24}$$

since these numbers correspond to the values of $\varphi_1'$, $\varphi_2'$ and $\varphi_3'$ at t=0, when these functions have their domains extended that far. We desire to compute

$$\Delta L_{1,f} := \int L_{1,f}(\varphi_U) d^4x - \int L_{1,f}(\varphi_L) d^4x .$$

Using Eq.3.23 we find that $\Delta L_{1,f}$ can be expressed as

$$\Delta L_{1,f} = \kappa_1 V_0 \{(\beta_1^2 - \beta^2)(t_2 - t_1) + (\beta_2^2 - \beta^2)(t_3 - t_2)\} . \qquad \text{Eq.3.25}$$

The endpoint conditions satisfied by $\varphi_1$, $\varphi_2$ and $\varphi_3$ imply that

$\alpha_1 \beta_1 \exp(\beta_1 t_1) = \alpha \beta \exp(\beta t_1)$, $\alpha_1 \beta_1 \exp(\beta_1 t_2) = \alpha_2 \beta_2 \exp(\beta_2 t_2)$, and $\alpha_2 \beta_2 \exp(\beta_2 t_3) = \alpha \beta \exp(\beta t_3)$,

and hence

$(\beta_1-\beta)t_1=\ln(\alpha\beta/(\alpha_1\beta_1))$, $(\beta_1-\beta_2)t_2=\ln(\alpha_2\beta_2/(\alpha_1\beta_1))$ and $(\beta_2-\beta)t_3=\ln(\alpha\beta/(\alpha_2\beta_2))$. Eq.3.26

Eq.3.26 permits us to write

$$t_2-t_1 = (\beta_1-\beta_2)^{-1}\ln(\alpha_2\beta_2/(\alpha_1\beta_1)) - (\beta_1-\beta)^{-1}\ln(\alpha\beta/(\alpha_1\beta_1)) \qquad \text{Eq.3.27}$$

and

$$t_3-t_2 = (\beta_2-\beta)^{-1}\ln(\alpha\beta/(\alpha_2\beta_2)) - (\beta_1-\beta_2)^{-1}\ln(\alpha_2\beta_2/(\alpha_1\beta_1)) . \qquad \text{Eq.3.28}$$

Upon combining Eqs.3.25, 3.27 and 3.28 we discover, after some effort, that

$$\Delta L_{1,f} = \kappa_1 V_0\{\beta\ln(\alpha_1\beta_1/(\alpha_2\beta_2)) + \beta_1\ln(\alpha_2\beta_2/(\alpha\beta)) + \beta_2\ln(\alpha\beta/(\alpha_1\beta_1))\} . \qquad \text{Eq.3.29}$$

To see that the right-hand side of Eq.3.29 is positive we note that since $0<t_3-t_2$ Eq.3.28 implies

$$0< (\beta_2-\beta)^{-1}\ln(\alpha\beta/(\alpha_2\beta_2)) - (\beta_1-\beta_2)^{-1}\ln(\alpha_2\beta_2/(\alpha_1\beta_1)) . \qquad \text{Eq.3.30}$$

By our construction we know that $\beta_2<\beta<\beta_1$. Hence $\beta_2-\beta<0$ and $\beta_2-\beta_1<0$. So if we multiply Eq.3.30 by $(\beta_2-\beta)(\beta_2-\beta_1)>0$ we get

$$0 < (\beta_2-\beta_1)\ln(\alpha\beta/(\alpha_2\beta_2)) + (\beta_2-\beta)\ln(\alpha_2\beta_2/(\alpha_1\beta_1)).$$

Simplification of this equation shows why the right-hand side of Eq.3.29 is indeed positive, since $\kappa_1>0$. Thus in this case we have demonstrated that the action of $L_{1,f}$ is indeed larger for $\varphi_U$ than it is for $\varphi_L$. In a similar way it can be shown that the value of the action of $L_{1,f}$ is always larger for the broken exponential path of a bump or dip then it is for the unbroken exponential path, regardless of whether the unbroken path is an increasing or a decreasing exponential section. (It is like an exponential version of the law of cosines for the action of $L_{1,f}$.) Consequently whenever a bump or dip is added to a class $C^{\infty,1}$ solution to the FLRW Cofinsler theories we have been investigating it increases the value of the action associated with the Lagrangian $L_{1,f}$ or $L_{1,f(k)}$. This clearly also applies when the Lagrangian of interest is the total Lagrangian

$$L_{T,1,f(k)} := L_{1,f(k)} + L_{f(k)}, \qquad \text{Eq.3.31}$$

since the action of $L_{f(k)}$ for the exponential solutions that begin and end with $\varphi=t$ is always 0. However, it must be noted that when using $L_{T,1,f(k)}$, $\varphi=t$, and more generally $\varphi=\gamma(\varepsilon_1 t+\varepsilon_2)^q$, are no longer solutions to the field equations. Although there are other solutions to these field equations than just $\varphi=\alpha e^{\beta t}$, they will not be required.

What the above analysis has shown us is that if we are interested in spaces that begin and end with $\varphi=t$, minimize the value of the action, $I(L_{T,1,f(k)})$, between $t=0$ and $t=t_c$, and are built from exponential solutions, it suffices to only consider universes which have two exponential portions. Such universes are like "vacuum fluctuations" of the $t<0$ spaces. Any other universes can be obtained from these by adjoining bumps or dips suitably, which increase the value of the action. For the universe built from two exponential sections we have

$$\varphi_2 := \begin{cases} t, & t<0 \\ \beta_1^{-1}\exp(\beta_1 t), & 0\leq t<t_1 \\ \alpha_2\exp(\beta_2 t), & t_1\leq t<t_c \\ t, & t_c\leq t, \end{cases} \qquad \text{Eq.3.32}$$

where $(\beta_1,\alpha_2,\beta_2) \in \mathbf{P_3}$, with $\alpha_2\beta_2\exp(\beta_2 t_c)=1$. The usual calculations show that in this case

$$I(L_{T,1,f(k)}(\varphi_2)) = \kappa_1 V_k \beta_1 \ln\alpha_2\beta_2 = -\kappa_1 V_k \beta_1 \beta_2 t_c > 0 \, . \qquad \text{Eq.3.33}$$

This bump universe built from two exponential arcs (with $\beta_1>0$, and $\beta_2<0$), is, as mentioned earlier, essentially just a vacuum fluctuation, and unlike our universe. For the universe built from three exponential arcs we have

$$\varphi_3 := \begin{cases} t, & t<0 \\ \beta_1^{-1}\exp(\beta_1 t), & 0\leq t<t_1 \\ \alpha_2\exp(\beta_2 t), & t_1\leq t<t_2 \\ \alpha_3\exp(\beta_3 t), & t_2\leq t<t_c \\ t, & t_c\leq t \, . \end{cases} \qquad \text{Eq.3.34}$$

where $(\beta_1,\alpha_2,\beta_2,\alpha_3,\beta_3) \in \mathbf{P_5}$. It is a straightforward matter to show that

$$I(L_{T,1,f(k)}(\varphi_3)) = \kappa_1 V_k[(\beta_1-\beta_3)\ln(\alpha_2\beta_2) + \beta_2\ln(\alpha_3\beta_3)] > 0 \, , \qquad \text{Eq.3.35}$$

which can be made as close to zero as we wish, but not for parameters suitable to our universe. If this were to be a model of our universe we would want $\beta_1 \gg 0$, $t_1 \ll 1$, and $\beta_2>0$ but very close to 0. Since $\beta_3$ is necessarily negative, to get $I(L_{T,1,f(k)}(\varphi_3))$ close to zero, we would need $\beta_3$ close to $0^-$, which correspondingly would yield a long period of exponential decline. However, we do not really know if that is the case for our universe. Nevertheless the action that we would get using Eq.3.35 would not be close to zero. So perhaps minimizing the action does not yield all the physically interesting universes.

Given a value of the action $I(L_{T,1,f(k)}(\varphi_3))$ close to $0^+$, we could probably construct a universe from more than three exponential arcs that had the same value for its action.

So the value of the action does not determine how many arcs comprise the universe. But since adding bumps and dips to a universe increases the value of the action, we could conclude that this second universe would have to have a smaller value of the cutoff time $t_c$. In other words, adding bumps and dips to a universe with the requirement that the action remain constant, shortens that universe's lifetime.

There exists a second way to modify the original Lagrangian $L_{f(k)}$ to get a new Lagrangian which admits solutions to its Euler-Lagrange equations of the form $\varphi=\alpha e^{\beta t}$, and for which the new Lagrangian is proportional to $\beta^2$ without any time dependence. This can be done by introducing a third scalar field $\xi$ on M, where $\xi \sim \ell$. In the context of our FLRW Cofinsler theory $\xi=\xi(t)$. I take the Lagrangian for $\xi$ to have the form

$$L_\xi = -\tau g^{\frac{1}{2}} \varphi^a |\, f^* \,|^b g^{ij} \xi_{,i} \xi_{,j} \qquad \text{Eq.3.36}$$

where $\tau>0$ is a constant, and a, b are to be determined. The total Lagrangian for the f, $\varphi$, $\xi$ system is taken to be

$$L_{T,\xi,f(k)} := L_\xi + L_{f(k)}. \qquad \text{Eq.3.37}$$

Using Eq.2.11 a familiar calculation shows that

$$L_\xi = \tau[(1-kr^2)^{-\frac{1}{2}} r^2 \sin\theta] \varphi^a (\varphi')^{2b+3} (\xi')^2. \qquad \text{Eq.3.38}$$

Upon varying $\xi$ in $L_{T,\xi,f(k)}$ we find that its associated Euler-Lagrange equation is

$$\frac{d}{dt}(\varphi^a (\varphi')^{2b+3} \xi')) = 0, \qquad \text{Eq.3.39}$$

and hence
$$\xi' = \mu \varphi^{-a} (\varphi')^{-(2b+3)} \qquad \text{Eq.3.40}$$

where μ is a constant. Employing Eqs. 2.7, 2.8, 2.12, 2.13, 3.37 and 3.38 shows us that when we vary φ in $L_{T,\xi,f(k)}$ we obtain the following Euler-Lagrange equation

$$\kappa(\varphi')^{-1}\frac{d^2}{dt^2}(\varphi(\varphi')^{-2}\varphi'') + \tau a\varphi^{a-1}(\varphi')^{2b+3}(\xi')^2 - (2b+3)\tau\frac{d}{dt}(\varphi^a(\varphi')^{2b+2}(\xi')^2) = 0, \text{ Eq.3.41}$$

which, due to Eq.3.40 becomes

$$\kappa(\varphi')^{-1}\frac{d^2}{dt^2}(\varphi(\varphi')^{-2}\varphi'') + \tau\mu^2[a\varphi^{-(a+1)}(\varphi')^{-(2b+3)} - (2b+3)\frac{d}{dt}(\varphi^{-a}(\varphi')^{-(2b+4)})] = 0 . \text{ Eq.3.42}$$

From our previous work we know that when $\varphi=\alpha e^{\beta t}$, the first term on the left-hand side of Eq.3.42 vanishes. So for $\varphi = \alpha e^{\beta t}$ to be a solution to Eq.3.42 we require either

$$a = -(2b+3) . \qquad \text{Eq.3.43}$$

or $b=-2$. When $b=-2$, Eqs.3.38 and 3.40 imply that $L_\xi(\varphi=\alpha e^{\beta t}) \propto \beta^{\text{some power}}$ only if $a=1$, which is what Eq.3.43 gives us when $b=-2$. So we can just require that a and b satisfy Eq.3.43. In that case we see that Eqs.3.38 and 3.40 imply that

$$L_\xi(\varphi=\alpha e^{\beta t}) = \tau\mu^2[(1-kr^2)^{-½}r^2\sin\theta] \beta^a . \qquad \text{Eq.3.44}$$

Thus we shall take a=2 and b=$-5/2$, so that $L_\xi$ given by Eq. 3.36 becomes

$$L_\xi = -\tau g^{½}\varphi^2| f^* |^{-5/2}g^{ij}\xi_{,i}\xi_{,j} , \qquad \text{Eq.3.45}$$

where $\tau \sim \ell^{-4}$. When $\varphi=\alpha e^{\beta t}$, a=2 and b =$-5/2$ we can use Eq. 3.40 to show that

$$\xi = \mu\beta^2 t + \xi_0 \qquad \text{Eq.3.46}$$

where $\mu \sim \ell^2$ and $\xi_0$ is a constant with $\xi_0 \sim \ell^1$.

Employing Eqs.2.12, 3.17, 3.45 and 3.46 to evaluate $L_{T,\xi,f(k)}$ for this solution to the field equations to obtain

$$L_{T,\xi,f(k)}(\varphi,\xi) = [(1-kr^2)^{-\frac{1}{2}}r^2\sin\theta]\,(\tfrac{1}{2}\kappa\beta + \tau\mu^2\beta^2).  \qquad \text{Eq.3.47}$$

However, now, unlike the previous situation with $L_{T,1,f(k)}$, the value of the integration constant $\mu$ can change for each exponential segment of the universe under investigation, and I have to state suitable boundary conditions for $\xi$. I shall take these boundary conditions to be that $\mu$ (as defined by Eq.3.40, with $a=-(2b+3)=2$) does not change as we pass from one exponential segment to the next, and $\xi_0$ is chosen so that $\xi$ is continuous. Thus for the $\varphi$ solution presented in Eq.3.34 I shall take

$$\xi := \begin{cases} 0, & \text{if } t<0, \\ \mu\beta_1^2 t, & 0\leq t<t_1 \\ \mu\beta_2^2(t-t_1)+\mu\beta_1^2 t_1, & t_1\leq t<t_2 \\ \mu\beta_3^2(t-t_2)+\mu\beta_2^2(t_2-t_1)+\mu\beta_1^2 t_1, & t_2\leq t<t_c \\ \mu\beta_3^2(t_c-t_2)+\mu\beta_2^2(t_2-t_1)+\mu\beta_1^2 t_1, & t_c\leq t. \end{cases} \qquad \text{Eq.3.48}$$

Due to Eqs.3.39 and 3.41 we see that $\xi$ = constant is a solution to the Euler-Lagrange equations of $L_{T,\xi,f(k)}$ when $\varphi=\alpha e^{\beta t}$ and $\varphi=t$. Hence when $\xi$ is given by Eq.3.48, and $\varphi$ is given by Eq.3.34, the pair $\varphi$, $\xi$ is a solution to the field equations for all time, not just for $0<t<t_c$. Recall that this is not the case for the Lagrangian $L_{T,1,f(k)}$, when $\varphi$ is given by Eq.3.34, for then the field equations are satisfied only when $0<t<t_c$.

Using the value of $L_{T,\xi,f(k)}$ given in Eq.3.47 we find, in analogy to Eq.3.35, that when $\varphi$ and $\xi$ are given by Eq.3.34 and 3.48 then

$$I(L_{T,\xi,f(k)}) = \mu^2\tau V_k[(\beta_1-\beta_3)\ln(\alpha_2\beta_2) + \beta_2\ln(\alpha_3\beta_3)] > 0\,. \qquad \text{Eq.3.49}$$

One interesting aspect of the action presented in Eq.3.49 is that if we evaluate it for a universe that begins in a "Guthian" manner, with $\beta_1 \gg 0$, and $\beta_2 \approx 0^+$, then although the term in square brackets can be large, the end result can be made very small by choosing $\mu$ close to zero. Moreover, when $\mu$ is small, Eq.3.48 implies that during the time $t_1 < t < t_2$ when the universe is exponentially "coasting," $\xi$ is fairly constant with $\xi \approx \mu \beta_1^2 t_1$.

I find it very interesting that we require three scalar fields f, $\varphi$ and $\xi$, to construct universe models which minimize the action and are pieced together from sections in which the scale factor, $\varphi'$, is exponential, except when $t<0$, and $t>t_c$, where it equals 1. It is even more intriguing if we can regard $\varphi$ as representing dark energy, and $\xi$ as dark matter.

It is easy to modify our multiverses presented in Eq.3.14 to incorporate a third scalar field $\xi$ that satisfies the boundary conditions that I stated above. To that end I define

$$\mathbf{P_{5,1}} := \{((\beta_1,\alpha_2,\beta_2,\alpha_3,\beta_3),\mu) \in \mathbb{R}^5 \times \mathbb{R} \mid (\beta_1,\alpha_2,\beta_2,\alpha_3,\beta_3) \in \mathbf{P_5} \text{ and } \mu \geq 0\}, \quad \text{Eq.3.50}$$

and set

$$\mathbf{MV_{k>0}(P_{5,1})} := T(\mathbf{P_{5,1}}) \times S_k, \; \mathbf{MV_{k=0}(P_{5,1})} := T(\mathbf{P_{5,1}}) \times \mathbb{R}^3, \; \mathbf{MV_{k<0}(P_{5,1})} := T(\mathbf{P_{5,1}}) \times H_k, \quad \text{Eq.3.51}$$

where $H_k$ is the 3-dimensional hypersphere of constant curvature k. The scalar field and metric tensor are defined on these multiverses just as they were on the previous multiverses, but now the field equations are derived from $L_{T,\xi,f(k)}$ and $\xi$ is given by Eq.3.48.

I have been alluding to probabilistic notions at various times in this section. It is time for me to bring these ideas to the forefront, although they are quite speculative. I suspect that universes in a multiverse with smaller values of the action should be more likely to be poplulated by matter when time begins. So, in keeping with the formula Feynmann uses in his path integral approach to quantum mechanics, I suggest that we define the probability that a universe in one of the above multiverses, represented by the pair $(\varphi_A, \xi_A)$, with $A \in \mathbf{P_{5,1}}$ is populated by matter is given by

$$\mathbf{P}((\varphi_A, \xi_A)) := N_{5,1}^{-1} \exp[-\hbar^{-1} I(L_{T,\xi,f(k)}(\varphi_A, \xi_A))] \qquad \text{Eq.3.52}$$

where $N_{5,1}$ is a normalizing factor given by

$$N_{5,1} := \int \exp[-\hbar^{-1} I(L_{T,\xi,f(k)}(\varphi_A, \xi_A))] d^6 x \qquad \text{Eq.3.53}$$

and the integral is taken over the parameter space $\mathbf{P_{5,1}} \subset \mathbb{R}^6$, with $d^6 x$ corresponding to the usual volume element in $\mathbb{R}^6$. In general $N_{5,1}$ does not exist, but I include it because its presence is formally useful, as we shall see in just a moment.

I would like to make a few comments about the formula for $\mathbf{P}((\varphi_A, \xi_A))$ presented in Eq.3.52. I put a minus sign in the argument of the exponential function appearing in the definition of $\mathbf{P}(\varphi_A, \xi_A)$ to guarantee that since $I(L_{T,\xi,f(k)}) \geq 0$, universes with larger action values are less probable to be populated when time begins. That is also the reason I choose to take the exponential of $-\hbar^{-1}$I in Eq.3.52, since exponentiating a large negative number is a good way to get a small number for the probability.

Since $N_{5,1}$ does not exist you might think that the definition presented in Eq.3.52 is useless. However, it is not completely useless since it allows us to discuss the ratio of the number of particles in the universes described by the pair $(\varphi_A, \xi_A)$ and $(\varphi_B, \xi_B)$ as time begins. This ratio can be taken to be given by

$$\mathbf{R}((\varphi_A,\xi_A);(\varphi_B,\xi_B)) := \mathbf{P}((\varphi_A,\xi_A))/\mathbf{P}((\varphi_B,\xi_B)) \qquad \text{Eq.3.54}$$

which, due to Eq.3.52, becomes

$$\mathbf{R}((\varphi_A,\xi_A);(\varphi_B,\xi_B)) = \exp[\hbar^{-1}(I(L_{T,\xi,f(k)}(\varphi_B,\xi_B)) - I(L_{T,\xi,f(k)}(\varphi_A,\xi_A))] \ . \qquad \text{Eq.3.55}$$

For those interested, the value of $\mathbf{R}$ for pairs of universes in the multiverse $\mathbf{MV_k(P_{5,1})}$ can be evaluated using Eq.3.49.

For completeness I shall now devote the remainder of this section to a discussion of megamultiverses, which will be 4-dimensional manifolds. For our scalar-scalar FLRW field theories these manifolds will be the analog of Hilbert spaces in Quantum Mechanics. I shall also say a few words about time travel in multiverses.

To begin $\forall\, n \in \mathbb{N}$ let

$$\mathbf{D_n} :=$$

$\{\,(\beta_1,\alpha_2,\beta_2,\ldots,\alpha_{n+1},\beta_{n+1},\mu) \in \mathbb{R}^+ \times \mathbb{R}^{2n} \times \mathbb{R} \,|\, \exists\, t_c > t_n > \ldots > t_1 > 0$ with $\exp(\beta_1 t_1) = \alpha_2\beta_2\exp(\beta_2 t_1) > 1$, $\alpha_2\beta_2\exp(\beta_2 t_2) = \alpha_3\beta_3\exp(\beta_3 t_2) > 1,\ldots,\alpha_n\beta_n\exp(\beta_n t_n) = \alpha_{n+1}\beta_{n+1}\exp(\beta_{n+1} t_n) > 1$ and

$\alpha_{n+1}\beta_{n+1}\exp(\beta_{n+1}t_c) = 1$, $\alpha_1 = 1$, $\beta_i \neq \beta_{i+1}$, for $i=1,\ldots,n$; $\mu \geq 0\}$. Eq.3.56

The requirement that $\alpha_i\beta_i\exp(\beta_i t_i) > 1$, guarantees that the scale factor $\varphi' = \alpha_i\beta_i\exp(\beta_i t)$, $t_{i-1} \leq t < t_i$, is always greater than 1 until $t=t_c$. The condition that $\beta_i \neq \beta_{i+1}$ assures that $\varphi$ is discontinuous when $t=t_i$. $\forall\ (\beta_1,\alpha_2,\beta_2,\ldots,\alpha_{n+1},\beta_{n+1},\mu) \in \mathbf{D_n}$ I construct a solution to the Euler-Lagrange equations of $L_{T,\xi,f(k)}$ by setting

$$\varphi :=$$

$t, t<0; \beta_1^{-1}\exp(\beta_1 t), 0 \leq t < t_1; \ldots; \alpha_i\exp(\beta_i t), t_{i-1} \leq t < t_i; \ldots; \alpha_{n+1}\exp(\beta_{n+1} t), t_n \leq t < t_c; t, t \geq t_c,$ Eq.3.57

$$\xi :=$$

$0, t<0; \mu\beta_1^2 t, 0 \leq t < t_1;\ldots;\mu\beta_i^2(t-t_{i-1})+\mu\beta_{i-1}^2(t_{i-1}-t_{i-2})+\ldots+\mu\beta_1^2 t_1, t_{i-1} \leq t < t_i;\ldots;\mu\beta_{n+1}^2(t-t_n)+$

$+\mu\beta_n^2(t_n-t_{n-1})+\ldots+\mu\beta_1^2 t_1, t_n \leq t < t_c; \mu\beta_{n+1}^2(t_c-t_n)+\mu\beta_n^2(t_n-t_{n-1})+\ldots+\mu\beta_1^2 t_1, t_c \leq t$. Eq.3.58

This $\varphi,\xi$ solution has $\varphi$ being of class $C^{\infty,1}$ with n+2 discontinuities, while $\xi$ is continuous and piecewise linear.

Now that I have defined $\mathbf{D_n}$ and the $\varphi,\xi$ solutions to the field equations I can present the multiverses which will be employed to construct the the megamultiverse.

$\forall\ n \in \mathbb{N}$, I set

$\mathbf{MV_{k>0}(D_n)} := \mathbf{\mathit{T}(D_n)} \times S_k$, $\mathbf{MV_{k=0}(D_n)} := \mathbf{\mathit{T}(D_n)} \times \mathbb{R}^3$, $\mathbf{MV_{k<0}(D_n)} := \mathbf{\mathit{T}(D_n)} \times H_k$, Eq.3.59

where $\mathbf{\mathit{T}(D_n)}$ is the 1-dimensional time manifold in $\mathbb{R} \times \mathbb{R}^{2n+2}$ built in the manner described earlier in this section. Thus we see that $\forall n \in \mathbb{N}$

$\mathbf{MV_{k>0}(D_n)} \subset \mathbb{R}^{2n+3} \times \mathbb{R}^4 = \mathbb{R}^{2n+7}$ and $\mathbf{MV_{k\leq 0}(D_n)} \subset \mathbb{R}^{2n+6}$, Eq.3.60

where I view $S_k$ as embedded in $\mathbb{R}^4$ in the usual way. A pair of scalar fields φ and ξ can be defined on each of the above multiverses in a manner analogous to the way it was done for the multiverses presented in Eq.3.14. The scalar field φ in turn gives rise to a Lorentzian metric tensor on the multiverse in accordance with Eq.1.8, and this metric tensor is everywhere continuous and of class $C^\infty$ except at the points of discontinuity in φ. The important thing about the construction is that $\forall\ n \in \mathbb{N}$, when t<0, φ=t, and ξ=0. I shall call this portion of the multiverses presented in Eq.3.59 the "tail" of that 4-dimensional multiverse. Note that this tail is actually a 4-dimensional open submanifold of each multiverse. The fact that all of the multiverses have the same tail, is the key to the construction of the megamultiverse. But before I do that I would like to say a few words about time travel.

Say we are given two points $(t_1,p_1)$ and $(t_2,p_2)$ in $T(D_n)$, where $t_2 > t_1 > 0$. We embed $T(D_n)$ in $T(D_n) \times \mathbb{R}$ as the "0 slice," and build a "shunt" in $T(D_n) \times \mathbb{R}$ from the point $((t_1,p_1),1)$ to $((t_2,p_2),1)$. $T(D_n)$ with this shunt is the underlying manifold of a new time manifold $T(D_n)'$, with its $C^\infty$ structure defined in the obvious way. If we cross $T(D_n)'$ with $M_k$ ( where $M_k$ could be either $S_k$, $\mathbb{R}^3$ or $H_k$, depending upon the choice of k, which is fixed) we obtain a 4-dimensional manifold with a global time function in which it is possible to travel from the branch of $T(D_n) \times M_k$ containing $(t_1,p_1)$ to the branch containing $(t_2,p_2)$. The problem, which I do not know how to solve, is to extend φ and ξ to the shunted manifold, $T(D_n)'xM_k$, so that they link up properly (although this is easy to do if $p_1=p_2$). So time travel is relatively easy within a multiverse, but it is not easy to bring your baggage, φ and ξ, with you. Another thing to be noted is that this time travel is not "local" since you take all of $M_k$ with you on this shunt, and not just the neighborhood of some point $((t_1,p_1),m)$.

In passing I would like to mention that suppose a "smooth landing" could be arranged as parts of two universes in $T(D_n)'xM_k$ join together. Then it might be possible to use the formula presented in Eq.3.55 to determine how much matter leaves one universe and enters the other.

Now for the megamultiverse. For each fixed value of $k \in \mathbb{R}$, I define

$$\mathbf{MV_k} := \bigcup_{n \in \mathbb{N}} \mathbf{MV_k(D_n)} \ . \qquad \text{Eq.3.61}$$

From Eq.3.60 we see that $\mathbf{MV_k} \subset \bigcup_{n \in \mathbb{N}} \mathbb{R}^{2n+7}$ or $\bigcup_{n \in \mathbb{N}} \mathbb{R}^{2n+6}$. The important thing to note is that each of the spaces whose union comprises $\mathbf{MV_k}$ has the same tail upon which φ and ξ are defined in the same manner. Thus we can clearly glue the tails of all the spaces in $\mathbf{MV_k}$ together to make one connected 4-dimensional quotient manifold, the megamultiverse, to be denoted by $MMV_k$.

The construction of the scalar fields φ and ξ on each $\mathbf{MV_k(D_n)}$, given in Eqs. 3.57 and 3.58, give rise to scalar fields on $\mathbf{MV_k}$ which pass to $MMV_k$ since φ=t and ξ=0, on each of the tails. These projected scalar fields will also be denoted by φ and ξ. This scalar field φ endows $MMV_k$ with a Lorentzian metric tensor via Eqs.1.7 and 1.8, when k=0, or Eqs.1.4 and 2.11 (for k≠0), with f defined on $T^*MMV_k$ in the expected manner.

These megamultiveses contain all of the possible universes that can be built from the exponential solutions for φ to the Euler-Lagrange equations derivable from $L_{T,\xi,f(k)}$, subject to the boundary conditions I have imposed on φ and ξ. But, as I mentioned earlier, there exists some non-exponential solutions to these Euler-Lagrange equations which might warrant investigation, and would lead to even bigger megamultiverses.

The notions of probability introduced earlier can be extended to $MMV_k$ as follows. If the pair of scalar fields $(\varphi_A, \xi_A)$ is defined on a branch of $MMV_k$, then $(\varphi_A, \xi_A)$ are also defined (as in Eqs.3.57 and 3.58) on the branch of some $MMV_k(D_n)$, with $A \in D_n$. I define the probability of the branch characterized by $(\varphi_A, \xi_A)$ to be populated as time begins at t=0 to be

$$\mathbf{P}((\varphi_A, \xi_A)) := N^{-1} \exp(-\hbar^{-1} I(L_{T,\xi,f(k)}(\varphi_A, \xi_A))), \qquad \text{Eq.3.62}$$

where

$$N := \sum_{i \in \mathbb{N}} \{ \int_{\mathbf{D_i}} \exp(-\hbar^{-1} I(L_{T,\xi,f(k)}(\varphi_A, \xi_A))) dx^{2i+2} \} \qquad \text{Eq.3.63}$$

and each integral is taken over the parameter space $\mathbf{D_i} \subset \mathbb{R}^{2i+2}$. Of course, each term in the expression for N is infinite, because of the "classical" range that the parameters in $\mathbf{D_i}$. I suspect that with suitable restrictions on the range of these parameters, N could be finite.

Lastly, if $(\varphi_A,\xi_A)$ and $(\varphi_B,\xi_B)$ characterize two different universes in $MMV_k$, then the ratio of the number of particles in the $(\varphi_A,\xi_A)$ universe to number of particles in the $(\varphi_B,\xi_B)$ universe when time begins should be

$$\mathbf{R}((\varphi_A,\xi_A);(\varphi_B,\xi_B)) := \mathbf{P}((\varphi_A,\xi_A))/\mathbf{P}((\varphi_B,\xi_B))$$

$$= \exp\{\hbar^{-1}[I(L_{T,\xi,f(k)}((\varphi_B,\xi_B))) - I(L_{T,\xi,f(k)}((\varphi_A,\xi_A)))]\}. \quad \text{Eq.3.64}$$

**Section 4: Concluding Remarks**

In this paper I have introduced a new type of geometry called Lorentzian Cofinsler Geometry, along with scalar-scalar theories of gravitation. There were basically two reasons behind my introduction of these notions. One was to enable the scalar field $\varphi$ to play a more active part in the construction of the geometry of spacetime, and in the process develop a means for generating the metric tensor from scalar potentials, analogous to the way the electromagnetic field is generated from a vector potential. In the case of gravity I required two potentials, the Cofinsler function, f, and the scalar field $\varphi$ to generate the metric tensor. My second reason for introducing this theory was the hope that it would simplify the equations of gravitation theory, and make them more amenable to machinery developed in the theory of classical mechanics. To demonstrate how these ideas can be employed we explored the cosmological solutions to the Euler-Lagrange equations of a particulary simple scalar-scalar Lagrangian given in Eqs.2.3 and 2.4. These solutions led us to consider the possibility that time can multifurcate ( like a tree developing an infinite number of branches from one limb), and in the process led us from one initial universe (or branch) to a myriad of other universes (branches); i.e., a multiverse. I tried to use the action associated with our Lagrangians to devise a method of determining which universes in our various multiverses are most likely to serve as a venue for particles once they are introduced. During the course of investigating these actions I was led to introduce another scalar field, $\xi$, which proved necessary to minimize the action for various physically plausible universes, such as our own.

Throughout this paper I never really discussed the "mouse in the room," namely (ordinary) matter, which approximately represents a mere 5% of the energy in our universe. Perhaps matter can be introduced as a perturbation of the other scalar fields, or through a matter Lagrangian that is added to $L_{T,\xi,f(k)}$. Or maybe matter locally satisfies Einstein's Equations (or some scalar-tensor modification thereof), with the metric tensor generated from f and $\varphi$ serving as asymptotic boundary conditions.

When I discussed time travel between various branches of a multiverse we saw that this time travel was of a global nature that would actually rip apart a branch. It might turn out that when matter is introduced time travel can become localized, because matter has a way of isolating portions of the universe through black holes. Maybe these isolated regions, in different branches, can be linked together so that the φ and ξ fields in these branches mesh together.

I hope that the material I have presented here has aroused your curiosity to explore alternative geometric cosmological theories, and in particular the possible role Cofinsler structures can play in that investigation.

**Acknowledgements**

I would like to thank Professors Spiros Cotsakis and Alexander Yefremov for inviting me to participate in this special issue of the Philosophical Transactions A of the Royal Society. It is truly an honor to present ideas I have been contemplating for many years along with the work so many distinguished physicists.

I would also like to thank University of Waterloo Professor Emeritus John Wainwright for convincing the Department of Applied Math, and the University of Waterloo, to appoint me to an Adjunct Professorship.

**Bibliography**

[1]     P. Finsler, "Über Kurven und Flächen in Allgemeinen Räumen," Ph.D. thesis, University of Göttingen, Göttingen, Germany, 1918.

[2]     H. Rund, "The Differential Geometry of Finsler Spaces," Springer-Verlag, 1959.

[3]  S.-S. Chern, D. Bao & Z. Shen, "An Introduction to Riemannian-Finsler Geometry," Springer-Verlag, 2000.

[4]  A. Bejancu & H. R. Farran, "Geometry of Pseudo-Finsler Submanifolds," Springer Science & Business Media B.V., 2000.

[5]  A. Friedmann, "Über die Krümmung des Raumes," Zeit. für Phys. A, **10**(1), 1922, 377-386. (doi: 10.1007/BF01332580)

[6]  G. Lemaître, "Expansion of the universe, A homogeneous universe of constant mass and increasing radius accounting for the radial velocity of extra-galactic nebulae," Mon. Not. R. Astron. Soc., **91** (5) 1931, 483-490.   (doi: 10.1093/mnras/91.5.483)

[7]  H.P. Robertson, "Kinematics and world structure," Astro. J. **82** 1935, 284-301. (doi: 10.1086/143681)

[8]  A.G.Walker, "On Milne's theory of world structure," Proc. London Math. Soc. Series 2, **42**(1) 1937, 90-127. (doi: 10.1112/plms/s2-42.1.90)

[9]  G. W. Horndeski, "Reformulating Scalar-Tensor Field Theories as Scalar-Scalar Theories Using Lorentzian Cofinsler Spaces," arXiv.org/abs/1911.07341, September, 2020.

[10] K. Schwarzschild, "Über das Gravitionsfeld eines Massenpunktes nach der Einsteinschen Theorie," Sitzungsber. Preuss. Akad. Wiss. **7** (1916), 189-196,

*c.f.*, arXiv:physics/9905030

[11] G. W. Horndeski, "Invariant Variational Principles and Field Theories," Ph.D. thesis, University of Waterloo, Waterloo, Ontario, 1973.

[12] G. W. Horndeski, "Second-Order Scalar-Tensor Field Theories in a Four-Dimensional Space," Inter. J. Theo. Phys. **10** (1974), 363-384. (doi: 10.1007/BF01807638)

[13] C.Deffayet, X. Gao, D. A. Steer & G. Zahariade, "From k-essence to generalized Galileons," Phys. Rev. D **84**, 064039 (2011); arXiv.org/abs/1103.3260, March, 2011. (doi: 10.1103/PhysRevD.84.064039)

[14] T. Kobayashi, M. Yamaguchi & J. Yokoyama, "Generalized G-inflation: Inflation with the most general second-order field equations," Prog. Theor. Phys. **126**, (2011) 511-526; arXiv.org/abs/1105.5723, September, 2011. (doi: 10.1143/PTP.126511)

[15] M. Ostrogradsky, "Memories sur les equations differentielles relatives au problems des isoperimetrics," Mem. Ac. St. Petersburg **6**, (1850), 385.

[16] R. P. Woodard, "The Theorem of Ostrogradsky," arXiv.org/abs/1506.02210, August, 2015.

[17] W. de Sitter, "On Einstein's Theory of Gravitation and its Astronomical Consequences," Mon. Not. R. Astron, Soc. **78** (1917), 3-28. (doi: 10/1093/

mnras/78.1.3)

[18] A. H. Guth, "The Inflationary Universe: The Quest for a New Theory of Cosmic Origins," Basic Books, 1998.

[19] F. Brickell. & R.S. Clark, "Differentiable Manifolds: An Introduction," Van Nostrand Reinhold Company, 1970.

[20] A.Linde, "The Eternally Existing, Self-Reproducing Inflationary Universe," Phys. Scr.**T15** (1987), 169-175. (doi: 10.1088/0031-8949/1987/T15/024)